\documentclass{osa-article}

\journal{osajournal}
\articletype{Research Article}
\def\@journalname{Applied Optics}
\definecolor{authorcolor}{RGB}{33,64,154} 
\definecolor{urlblue}{RGB}{46,46,177}

\usepackage{lineno}
\usepackage{subfigure}
\begin{document}
\title{Practical underwater quantum key distribution based on decoy-state BB84 protocol}

\author
{
SHANCHUAN DONG,\authormark{1} 
YONGHE YU,\authormark{1} 
SHANGSHUAI ZHENG,\authormark{1}
QIMING ZHU,\authormark{1}
LEI GAI,\authormark{1}
WENDONG LI,\authormark{1,2}
and YONGJIAN GU\authormark{1,*}
}
\address
{
\authormark{1}College of Physics and Optoelectronic Engineering, Ocean University of China, Qingdao 266100, China.\\
\authormark{2}liwd@ouc.edu.cn
}
\email{\authormark{*}yjgu@ouc.edu.cn}


\begin{abstract}
Polarization encoding quantum key distribution has been proven to be a reliable method to build a secure communication system. It has already been used in inter-city fiber channel and near-earth atmosphere channel, leaving underwater channel the last barrier to conquer. Here we demonstrate a decoy-state BB84 quantum key distribution system over a water channel with a compact system design for future experiments in the ocean. In the system, a multiple-intensity modulated laser module is designed to produce the light pulses of quantum states, including signal state, decoy state and vacuum state. The classical communication and synchronization are realized by wireless optical transmission. Multiple filtering techniques and wavelength division multiplexing are further used to avoid crosstalk of different light. We test the performance of the system and obtain a final key rate of 245.6 bps with an average QBER of 1.91\% over a 2.4m water channel, in which the channel attenuation is 16.35dB. Numerical simulation shows that the system can tolerate up to 21.7dB total channel loss and can still generate secure keys in 277.9m Jelov type  \uppercase\expandafter{\romannumeral1} ocean channel.
\end{abstract}
\section{Introduction}
The concept of quantum key distribution(QKD) and the first QKD protocol were proposed by Bennett and Brassard in 1984, known as the BB84 protocol\cite{bennett1984proceedings}. Its unconditional security in theory shine a light on a brand new territory of cryptography. Ever since, numerous studies have been performed based on the BB84 protocol in theory and experiments\cite{lo1999unconditional,shor2000simple,bruss1998optimal,inoue2002differential,gottesman2004security}. Notably there are the B92 protocol\cite{bennett1992quantum} and the decoy-state protocol\cite{lo2005decoy,wang2005beating,ma2005practical},which intend to simplify the system structure or to improve the system security, respectively. Experimentally, researchers have already built the experimental quantum channel to perform the space-to-ground secure communication over a range of 1200 km\cite{yin2017satellite,liao2017satellite},  Meanwhile, quantum channels that allow intercity secure communications have also been established through optical cables of a few hundred kilometers \cite{yin2016measurement,boaron2018secure}. Yet, there is still one piece of missing element for the global quantum networks, namely the underwater unconditionally secure wireless communication system.

Combining the wireless optical communication with quantum techniques will allow underwater vehicles to achieve high-speed and secure information exchanges. However, a big challenge is the huge attenuation of water, and the original BB84 protocol can not sustain such a huge channel loss. It has been proved that the decoy-state BB84 protocol can preserve the unconditional security of QKD and tolerate up to 50dB of channel loss\cite{wang2013direct,nauerth2013air,fedrizzi2009high}, making it a promising way to tackle the high attenuation of water. On the other hand, researchers have proved experimentally that photonic polarization states and entangled states can be preserved in water channels of different scales \cite{ji2017towards,hu2019transmission,zhao2019experimental,li2019proof}. Another big challenge is that unlike the telecom wavelengths, there is no effective polarization modulator for the blue-green region in current technology. As a result, multiple lasers have to be used in polarization encoding QKD systems\cite{liao2017satellite,hu2021decoy,yu2021experimental} . One of the latest experimental achievement\cite{hu2021decoy} is that a decoy-state QKD system has successfully generated secret keys through 30m Jerlov types \uppercase\expandafter{\romannumeral2}C seawater. However, the system still requires the high performance of arbitrary wave generators, local networks and other facilities to complete the key distribution procedure. Meanwhile, our team has also made a progress in developing practical underwater QKD systems \cite{yu2021experimental}. We built an all-optical polarization encoding decoy-state BB84 QKD system with complete electronic control, successfully generating secure keys through a 10.4m Jerlov type \uppercase\expandafter{\romannumeral3} seawater channel. However, the relatively large structure and complicate optical design are major obstacles to practical applications.

In this work, we successfully devise a new decoy-state underwater QKD system with compact structural design and complete system control. In order to achieve the purpose of compactness, we ditch the 8 quantum laser modules from our previous work and replace them with 4 sets of self-designed decoy-state laser modules. Each of the new laser module can produce light pulses of signal state, decoy state or vacuum state, which are required by the decoy-state method. Based on the 4-laser module scheme, we manage to simplify and renew the optical and structural design. The classical communication and system synchronization are both achieved by wireless optical transmission to guarantee practicality. In the lab experiments, we successfully perform the complete QKD process through water channel with the attenuation up to 16.35dB, which equals to 209m Jerlov type \uppercase\expandafter{\romannumeral1} ocean water. The numerical simulation results show the system can generate secure keys in 277.9m Jelov type I water channel, in which the total channel loss is as high as 21.7dB. This compact decoy-state QKD system with the complete FPGA (Field Programmable Gate Array) control allows us to move one step forward to practical underwater applications. 

\section{ System setup and methods}
We aim to devise a practical underwater decoy-state QKD system. The first priority is to keep the whole system as compact as possible. In 2017, researchers have successfully established the decoy-state QKD with polarization encoding from the Micius satellite to the ground over a distance of up to 1200km\cite{liao2017satellite}. The QKD transmitter on the satellite uses an 8-laserdiode scheme to implement the decoy-state BB84 protocol. In the previous work of our team\cite{yu2021experimental}, we also used 8 sets of laser modules to generate quantum light. But the relatively large system size made it difficult to move it or to integrate it into a water-tight platform to conduct marine tests. Therefore, it will be of practical importance to generate the decoy-state quantum light in a more compact way. In the following, we present a self-designed decoy-state laser module and a redesigned underwater decoy-state QKD system based on the 4-quantum-laser scheme to achieve this goal.

\subsection{Self-designed decoy-state laser module }
In order to carry out the decoy-state BB84 QKD protocol under a 4-quantum-laser scheme, we need every quantum laser module to generate optical pulses with three different intensities. According to this demand, we design and assemble a decoy-state laser module shown in Figs.1(a) and 1(b). For test and debug purposes, the module has yet to be integrated to a minimal size, which is approximately  $10cm\times12cm\times7cm$. Compared to the commercial laser module our team has used before, our current design has the advantage of size smallness. The laser diode (Osram PLT-450B) used in the laser module is controlled by a drive board. The drive board consists of a dual-channel laser switch chip (IC-Haus IC-HKB) and a related support circuit. The laser module is powered by an independent DC-DC power supply board based on the LM2596-ADJ voltage regulator. To keep the output power as stable as possible, a close-loop temperature control circuit is also integrated into the laser module.

This self-designed decoy-state laser module can be modulated by two independent channels. Here we define channel 1 (CH1) as the triggering signal state and channel 2(CH2) as the triggering decoy state. Both channels will deliver vacuum states after setting to low voltage levels. Additionally, the pulse intensity of the signal state and the decoy state can be separately adjusted and calibrated as demand. During the QKD experiments, the quantum trigger signal is sent by an FPGA board according to a 4-bit random number generated by the noise chip (WNG8), and the repetition frequency of the quantum trigger signal is 20MHz. 

 It is necessary to characterize the performance of the decoy-state laser module before the system integration. Figs. 1(c) and 1(d) present the temporal shape and the spectra of the optical pulses. Both of the test results are obtained under a modulation signal of 50MHz, 10ns pulse width. The full width at half maximum of the optical pulses triggered by the two channels is both less than 10ns. The spectra of the two optical pulses also show high consistency. To test the stability of the optical power, we use the optical power monitor (Thorlabs PM121D) to log the power data. The optical power variations under the two channels’ modulation are both less than 0.9\% during 60 minutes. The overall test results show that our self-designed decoy-state laser module has reached the design specification and can be integrated into the decoy-state QKD system for subsequent experiments. 

\begin{figure}[htbp]
    \centering
    \includegraphics[width=6cm]{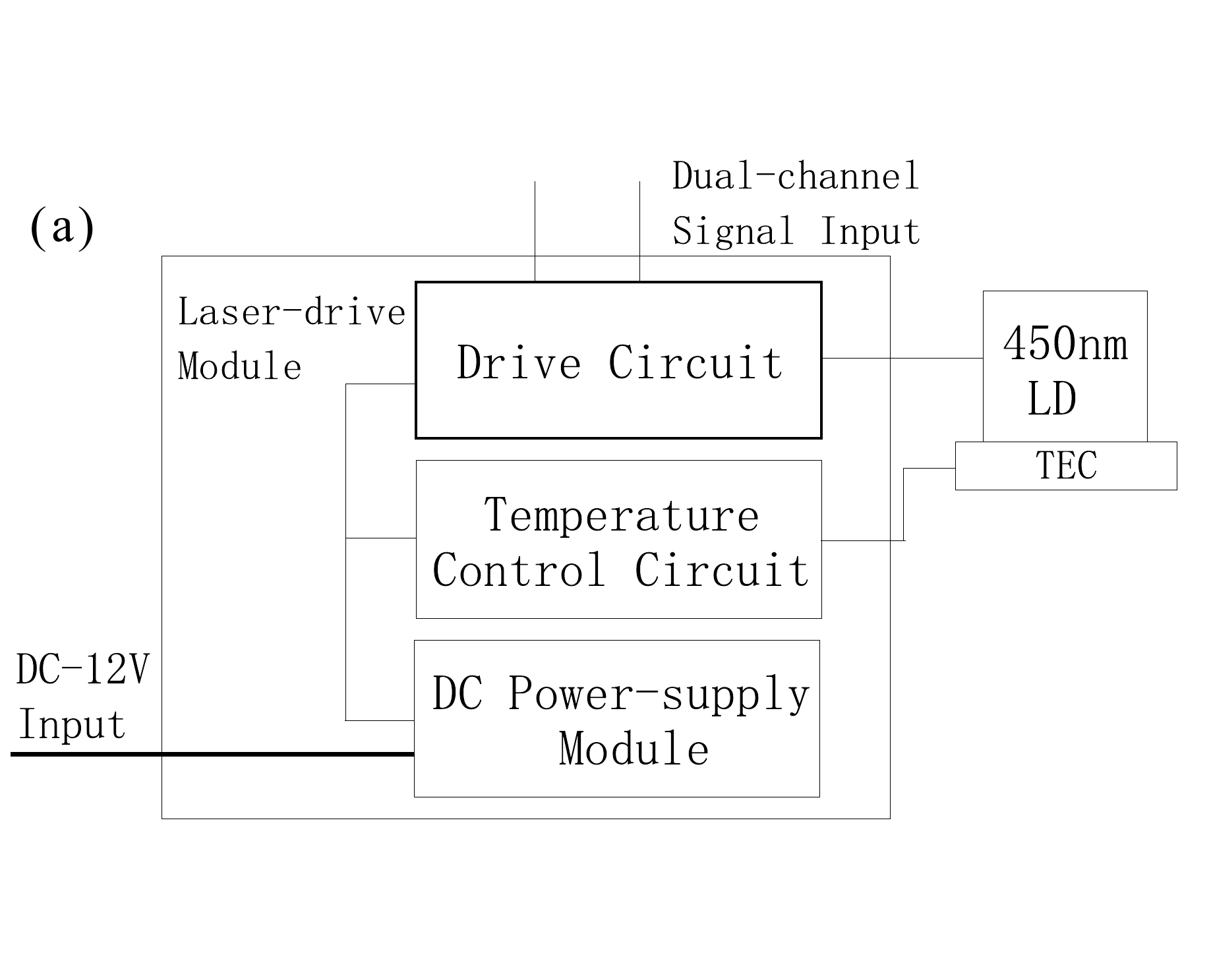}\includegraphics[width=6cm]{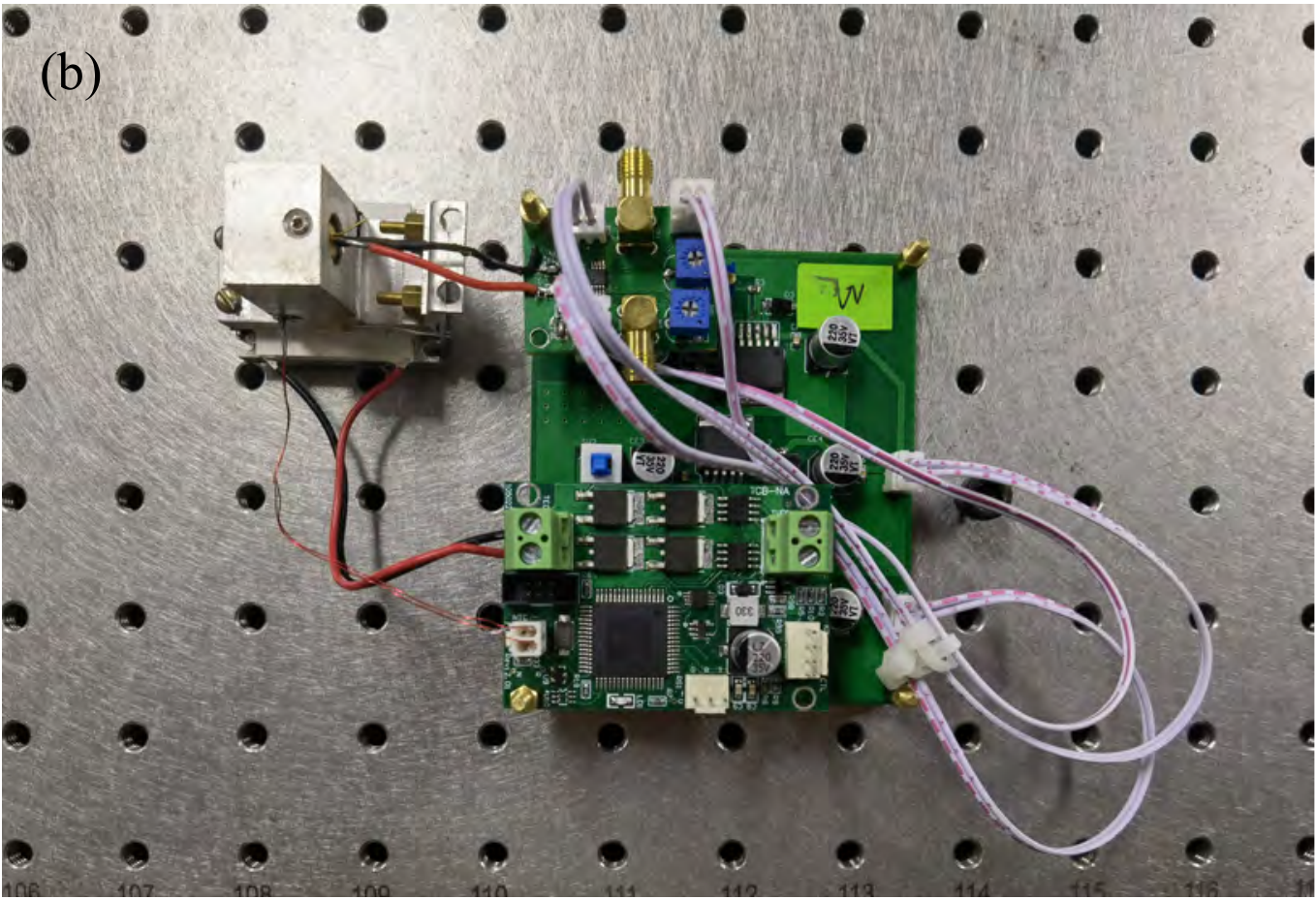}
    \includegraphics[width=6cm]{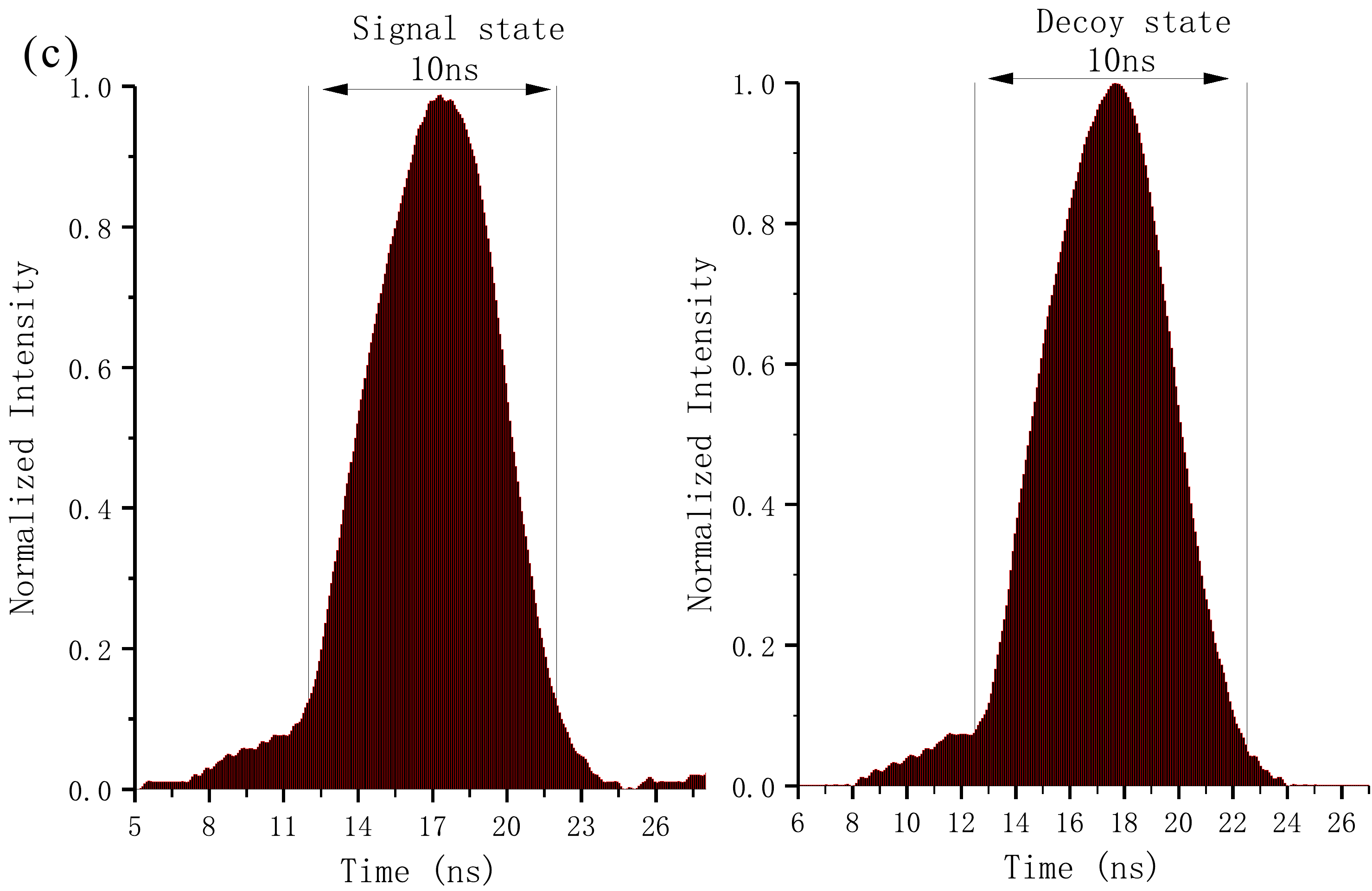}\includegraphics[width=6cm]{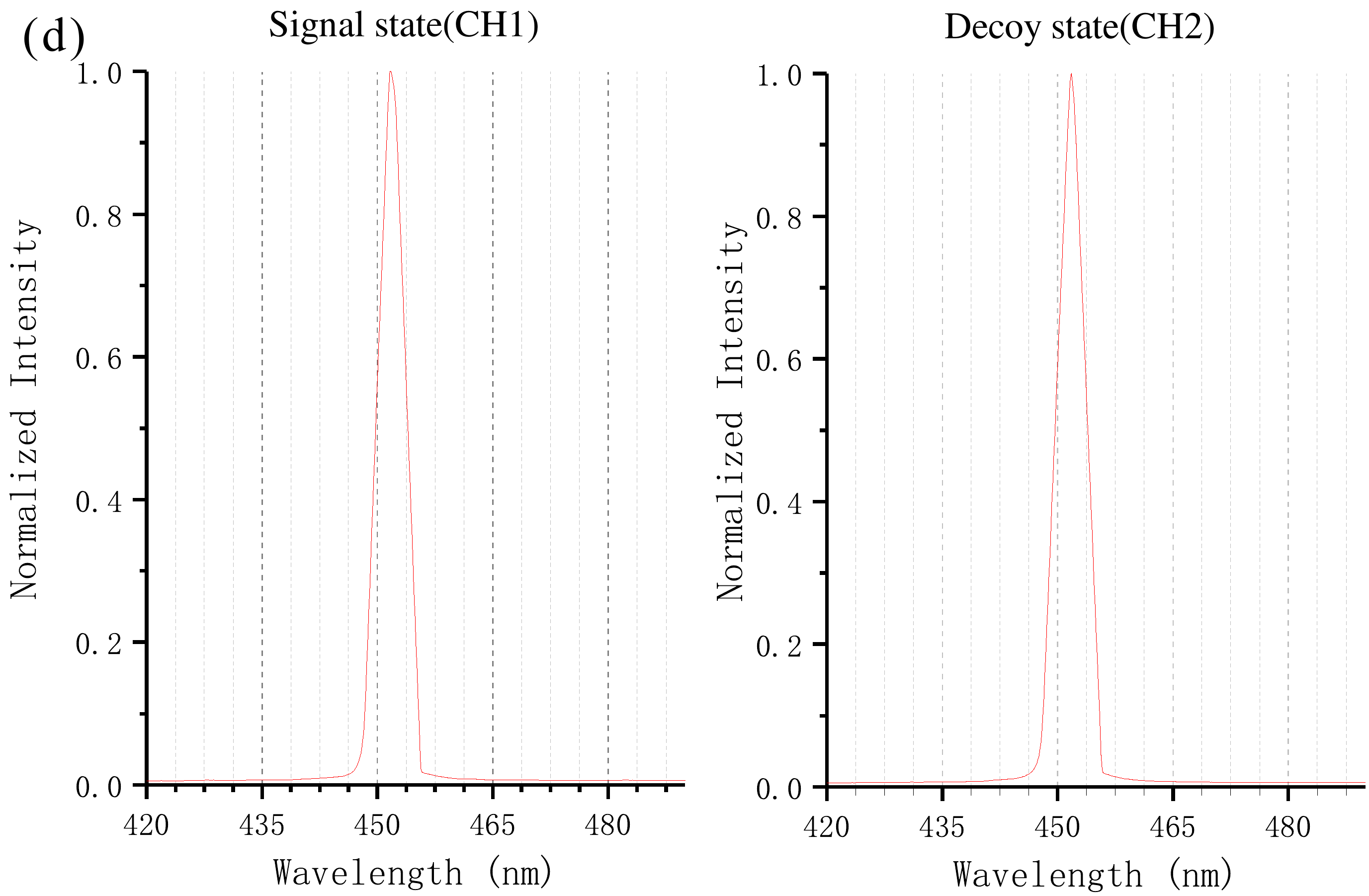}
    \caption{Self-designed decoy-state laser module. (a) and (b) show the schematic diagram and photograph of the self-designed decoy-state laser module. (c) Temporal shape of the signal state and decoy state optical pulses, with a modulation frequency 50MHz and 10ns pulse width. (d) Spectra of the signal state and the decoy state optical pulses.}
\end{figure}

\subsection{System layout}
Based on our four sets of self-designed decoy-state laser modules, we design and build a new compact decoy-state QKD system. The system setup is shown in Fig. 2. The Alice (transmitter) end and Bob (receiver) end of the system are separately mounted on two aluminum device decks. The optical components are fixed on each device deck using custom-made aluminum mounts. The dimension of each device deck is approximately $64.5cm\times23.5cm\times20cm$, significantly smaller than our previous designed system in Ref.\cite{yu2021experimental}.In underwater environments, only blue-green lights can sustain relatively long-distance propagations, while the laser commonly used in free-space air will decay rapidly. In this system, we use lasers with center wavelengths at 450nm, 520nm and 488nm for the quantum signal, classical signal and synchronization signal, respectively. Due to the lack of high-performance polarization modulators in the blue-green region, the four polarization states for implementing the BB84 protocol will be generated by optical components separately.

At the Alice end, the overall structure of the system is shown in Fig. 1(a). The quantum signals are generated by four 450nm self-designed decoy-state laser modules. They are combined into one beam by the beam-splitter (BS), polarization beam splitter (PBS) and mirrors after heavy attenuation. While the 488nm synchronization beam and the 520nm classical beam are combined into one beam using a dichroic mirror (DM). The two combined beams are then transmitted through the water channel to the Bob end. The quantum signals are encoded into four polarization states by half-wave plates and PBSs, corresponding to the horizontal (H), vertical (V), 45°(P), and 135°(M), respectively. Meanwhile, the three-intensity optical pulse under each polarization state required by the decoy-state method will be generated by a self-designed decoy-state laser module according to the modulation signal. 

At the Bob end, the structural design is shown in Fig. 1(b). The quantum signals sent by Alice will be detected by single photon detectors under four different polarization bases. The polarization basis is selected randomly by a 50:50 beam splitter, two polarization beam splitters and a half-wave plate. The two avalanche photodiode detectors (APDs) are used to detect the synchronization signal and the classical signal. Meanwhile, another 520nm laser module is used to accomplish the duplex classical communication with Alice.

\begin{figure}[htbp]
\centering\includegraphics[width=12cm]{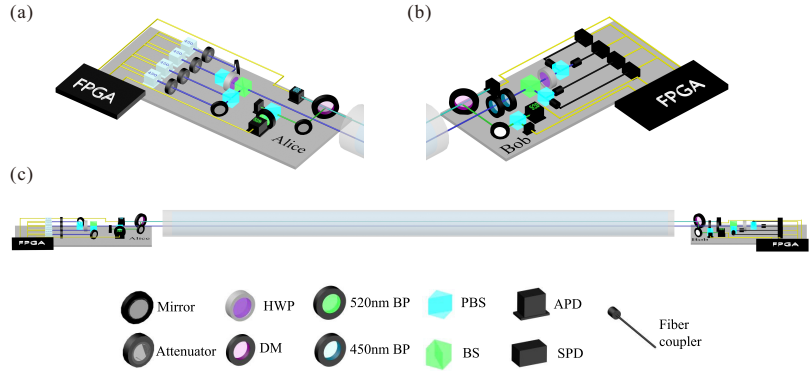}
\caption{System layout.(a) and (b) show schematic diagrams of Alice and Bob ends. (c) Experimental setup, the Alice end and the Bob end are separated by a 2.4m water channel. FPGA is used at both ends as system control units. HWP: half-wave-plate, DM: dichroic mirror, BP: bandpass filter, PBS: polarization beam splitter, BS: beam splitter, APD: avalanche photodiode (Hamamatsu C12702-11), SPD: single photon detector (Hamamatsu H12386).}
\end{figure}

\subsection{System control scheme}
A complete system control scheme is also crucial for a practical underwater decoy-state QKD system. We inherit the general idea of our previous work and use the FPGA board as the main control unit at both the Alice and Bob ends. At the Alice end, instead of a relatively large arbitrary wave generator, the quantum key signal is generated by FPGA and sent to the quantum laser module. The key sifting process is also accomplished within FPGA on both ends. The post-processing procedure including the error checking, error correction and privacy amplification is finished on users’ PC. 

For the system timing synchronization scheme, Alice will keep sending a 5MHz synchronization signal to Bob during QKD processes. The synchronization signal will then be used to generate a 20MHz gate signal in the  Bob’s FPGA. Meanwhile, a self-adjusting algorithm is used in FPGA to control the gate signal and to locate the position of the quantum signal in time domain. Therefore, the FPGA will not log the noise photon (outside the gate) as the quantum key and increase the quantum bit error rate. Moreover, the synchronization signal also serves as a reference clock, which helps to identify the right sequence of incoming quantum keys. As above, we are able to detect quantum signals with stable and correct time sequences under this synchronization scheme.

For the classical duplex communication, the information exchange between Alice and Bob is realized by two 520nm lasers on both ends. The FPGA will control the 520nm lasers by the open-off-keying modulation, with a signal baud rate of 20MHz. These two lasers will emit orthogonal polarization lights (H and V), so that the optical pulses of classical communication can share the same light path during the transmission and be easily separated by PBS on both ends. 

\subsection{ Filtering and wavelength division multiplexing }
During the QKD process, quantum signals are extremely weak at the receiver side. Therefore, other than the self-adjusting gate used to provide filtering in time domain, we also apply filtering techniques in both frequency and spatial domains. At the Alice end, we have 488nm and 520nm band pass filters (linewidth 5nm) placed in front of the synchronization laser and the classical laser. At the Bob end, we seal the quantum signal receiving components in a black box and place two 445nm bandpass filters (linewidth 20nm) at the entrance of quantum signals. These bandpass filters could minimize the effects of background noise and the crosstalk of classical lights (488nm and 520nm). In spatial domain, we use small caliber fiber couplers to gather the photons of quantum signals. Compared to directly using the SPD to gather photons in free space, the fiber coupler can restrict the field angle to around 0.14 degrees. This spatial filtering measure will significantly eliminate noise photon counts caused by the background light and classical signal.

Wavelength division multiplexing techniques have also been used in our system for reducing alignment difficulties. At the Alice end, a 496nm low-pass-high-reflect dichroic mirror is placed in front of the synchronization laser (488nm), where the 520m beam will be reflected and combined with the 488nm beam. At the Bob end, the joint beam will be split by another dichroic mirror and captured by corresponding detectors.

\section{ Experiments procedure and results}
Before conducting the QKD procedure, we use a 450nm laser (quantum signal) to characterize the channel loss. The channel we used is a 2.4m toughened glass tank filled with water. Since the parameter of the tank is fixed, we manage to adjust the water quality to simulate different channel conditions. Eventually we got four degrees of attenuation: 10.22dB, 12.36dB, 14.4dB and 16.35dB. the corresponding attenuation coefficients are 0.98m$^{-1}$, 1.187m$^{-1}$,1.383m$^{-1}$ and 1.569m$^{-1}$. 

The average water temperatures under all three conditions are maintained at 15$^{\circ}$C, while the fluctuation is below 0.1$^{\circ}$C. The water turbulence within the tank is not observed. The tank is sitting still on the optical table and no mechanical vibrations are introduced to the table. The optical attenuation is 4.1dB at the Bob end due to imperfect alignments and component losses. Taking the 20\% SPD detecting efficiency into account, the total optical attenuation at the Bob end is 11.1dB.

The decoy-state QKD protocol requires three different intensities of light pulses. By utilizing our self-designed decoy-state laser module, we can carefully calibrate the average photon number per pulse of the signal and decoy states with the assist of attenuators. Here we set the average photon number per pulse of the signal state to 0.8 and the decoy state to 0.1, while the photon number of vacuum state is zero. As we mentioned earlier, the quantum trigger signal is sent by the FPGA according to a 4-bit random number. This random number will determine the quantum state under which polarization will be emitted and the corresponding relation is shown in Tab. 1. Notably the general probabilities of the three states is 2:1:1.

\begin{table}[]
	\renewcommand\arraystretch{1.5}
	\caption{Corresponding relations of 4-bit random numbers and photon states}\label{Table 1}
    \centering  
    \scriptsize
	\begin{tabular}{|p{17pt}<{\centering}|p{17pt}<{\centering}|p{55pt}<{\centering}|p{17pt}<{\centering}|p{17pt}<{\centering}|p{75pt}<{\centering}|}
		\hline		
		$Bit_0$&$Bit_1$&$Quantum State$&$Bit_2$&$Bit_3$&$Polarization State$\\
		\hline
		0&0&$Vacuum State$&0&0&$H$\\
		\cline{1-6}\cline{2-6} \cline{3-6} \cline{4-6} \cline{5-6} \cline{6-6} 
		0&1&$Decoy State$&0&1&$V$\\
		\cline{1-6}\cline{2-6} \cline{3-6} \cline{4-6} \cline{5-6} \cline{6-6} 
		1&0&$Signal State$&1&0&$P$\\
		\cline{1-6}\cline{2-6} \cline{3-6} \cline{4-6} \cline{5-6} \cline{6-6} 
		1&1&$Signal State$&1&1&$M$\\
		\cline{1-6}\cline{2-6} \cline{3-6} \cline{4-6} \cline{5-6} \cline{6-6} 
		\hline
	\end{tabular}
	\normalsize
\end{table}  

After the calibration of the Alice’s quantum laser module, we send the four polarization states to Bob to verify the fidelity of these polarization states. The reconstructed density matrices of all polarization states and their corresponding fidelities are shown in Fig. 3. The average value of the polarization state fidelity is 98.574\%, which is fully capable of proceeding subsequent QKD procedures.

\begin{figure}[htbp]
    \centering
    \subfigure[]{\includegraphics[width=9cm]{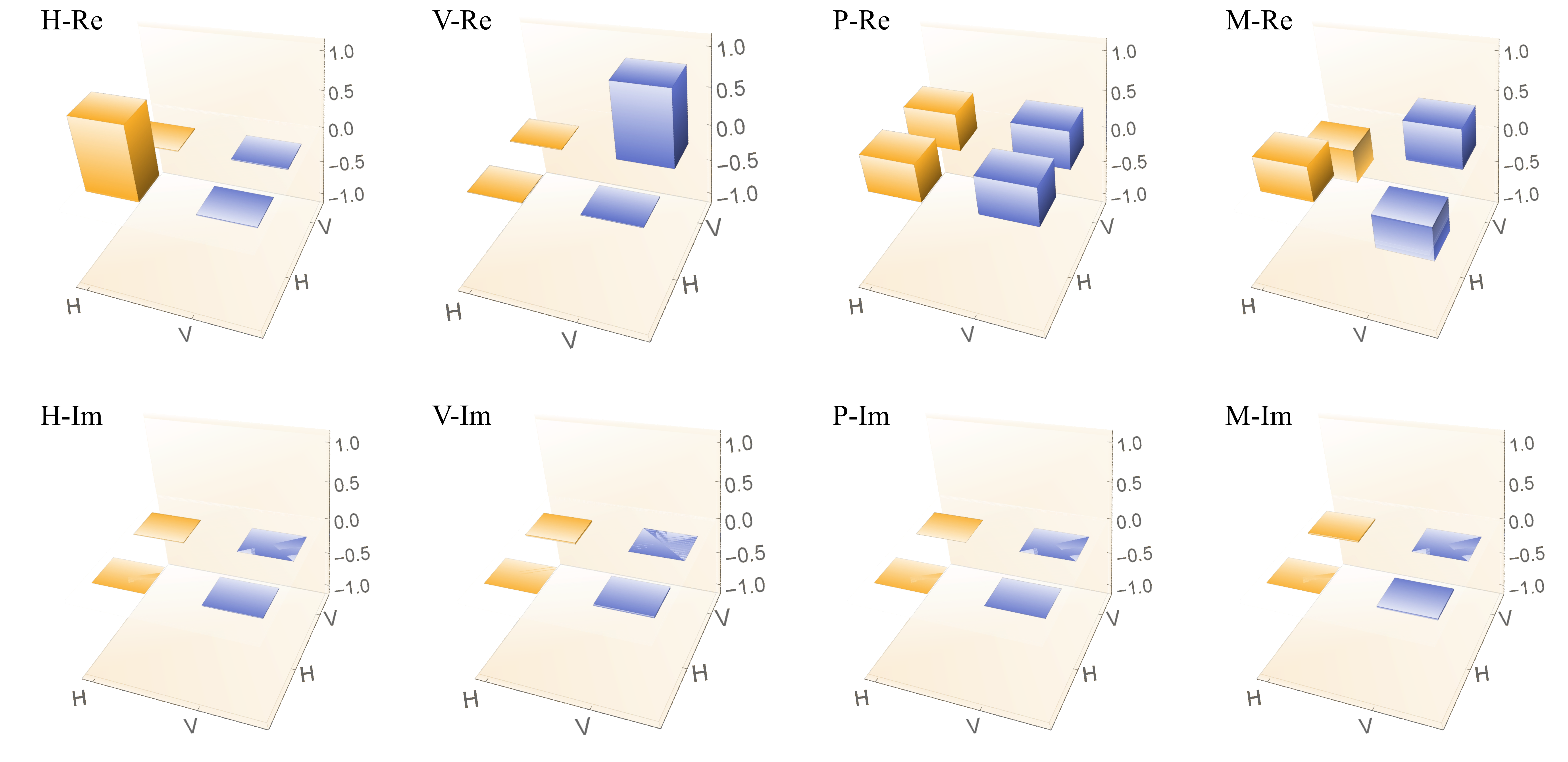}}
    \centering
    \subfigure[]{\includegraphics[width=7cm]{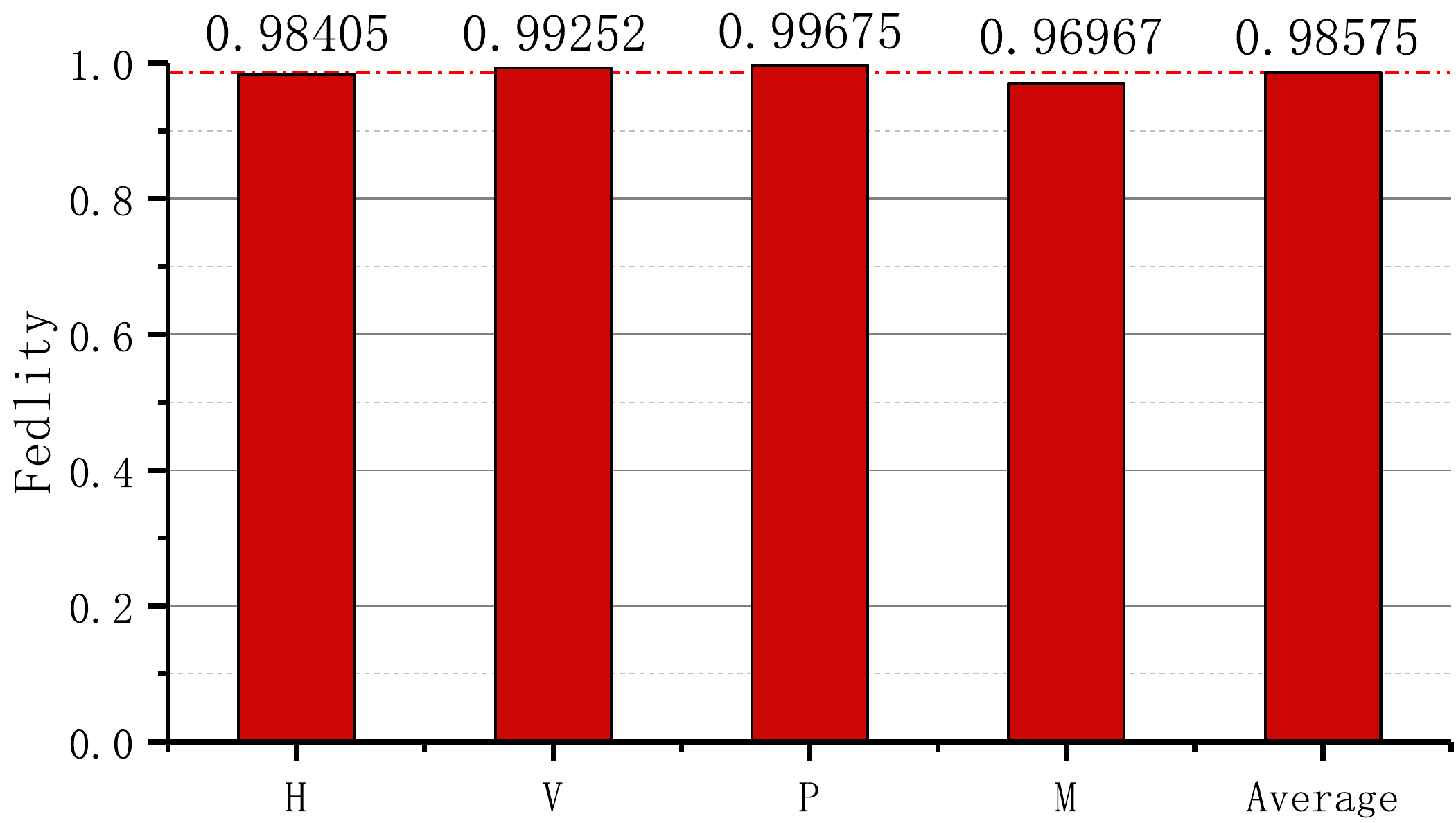}}  
    \caption{Fidelity analysis. (a) Reconstructed density matrices of different polarization states according to the experimental data. The real and imaginary parts are separately demonstrated. (b) The fidelity of each polarization state.}
\end{figure}

We continuously run the decoy-state QKD system under three different water channel attenuation. The gain of signal state and the bit error rate are shown in Fig. 4. The detailed parameters of the QKD experiments are shown in Tab. 2. Fig.4(a),(b) and (c) show the gain and bit error rate of signal states under different channel attenuation. For the 10.22dB case, our system has been stably and continuously run for over 520s. The average gain of signal state $Q_u$ and decoy state $Q_v$ have reached $1.48\times10^{-2}$ and $2.05\times10 ^{-3}$ respectively, while the average bit error rate are $E_u$1.21\% and $E_v$1.81\%. During the experiment, we obtain a average final key rate of 3535.7 bits/s. 

\begin{figure}[hbtp]
    \centering
    \includegraphics[width=6.5cm]{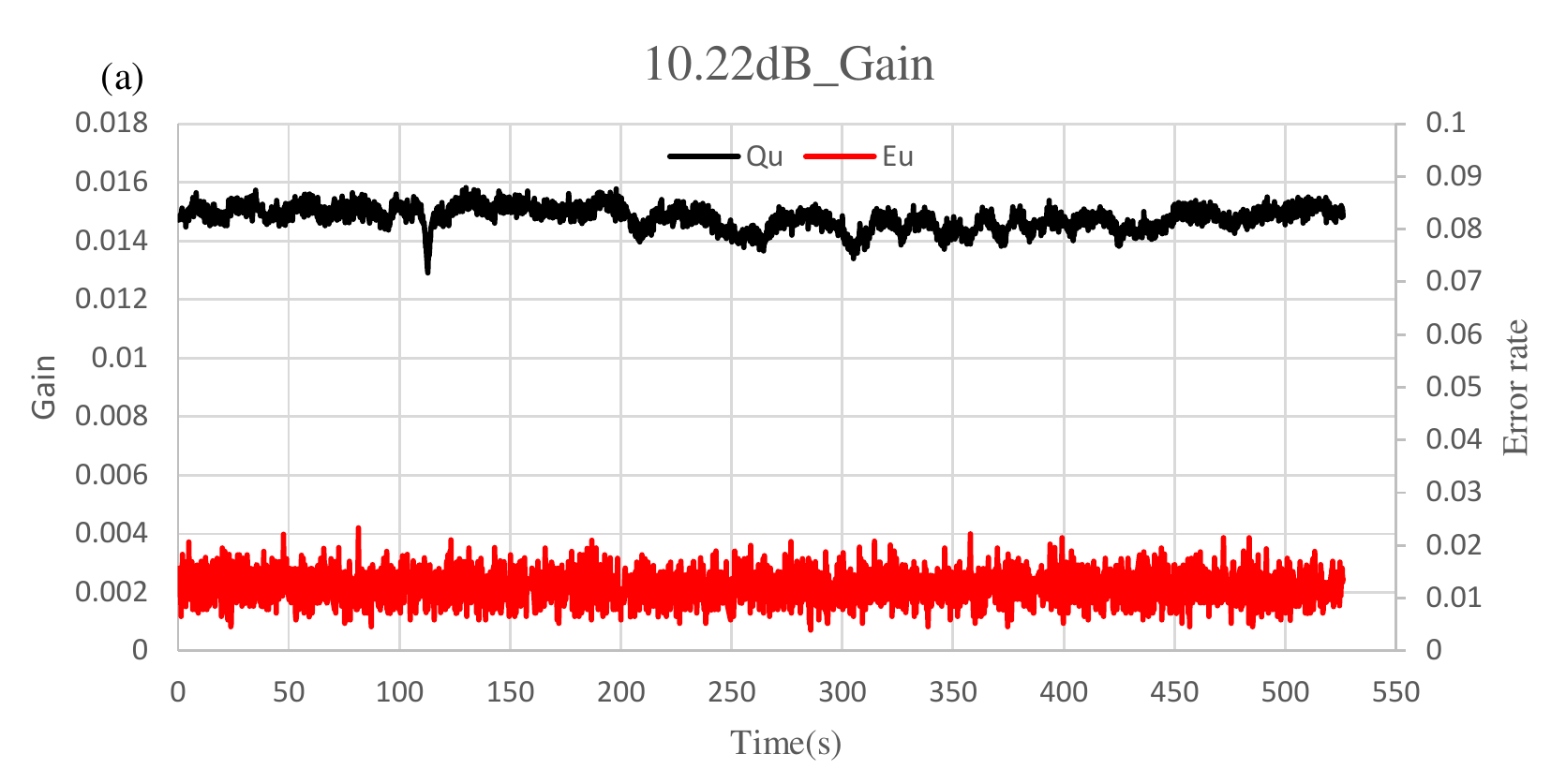}\includegraphics[width=6.5cm]{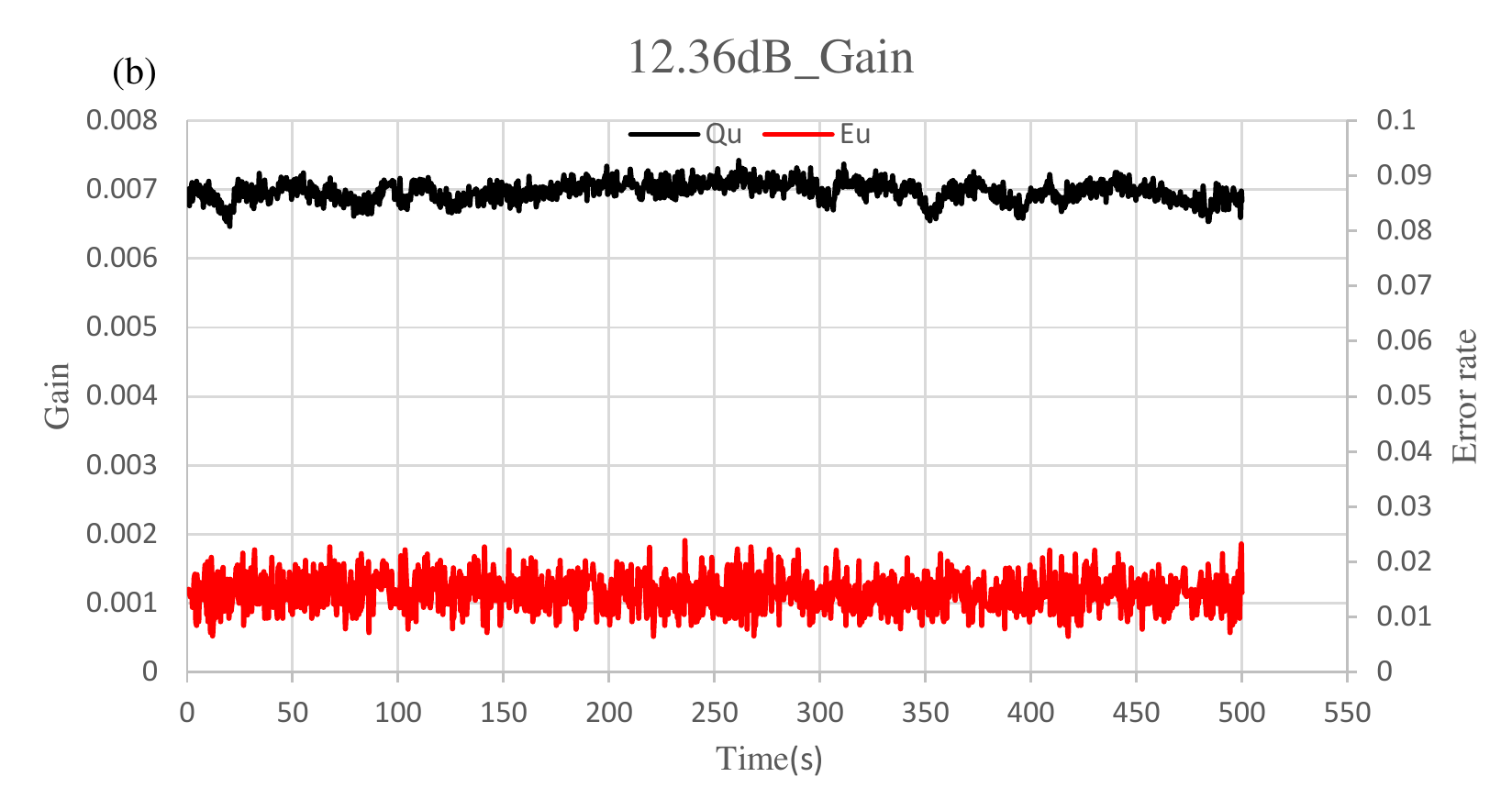}
    \includegraphics[width=6.5cm]{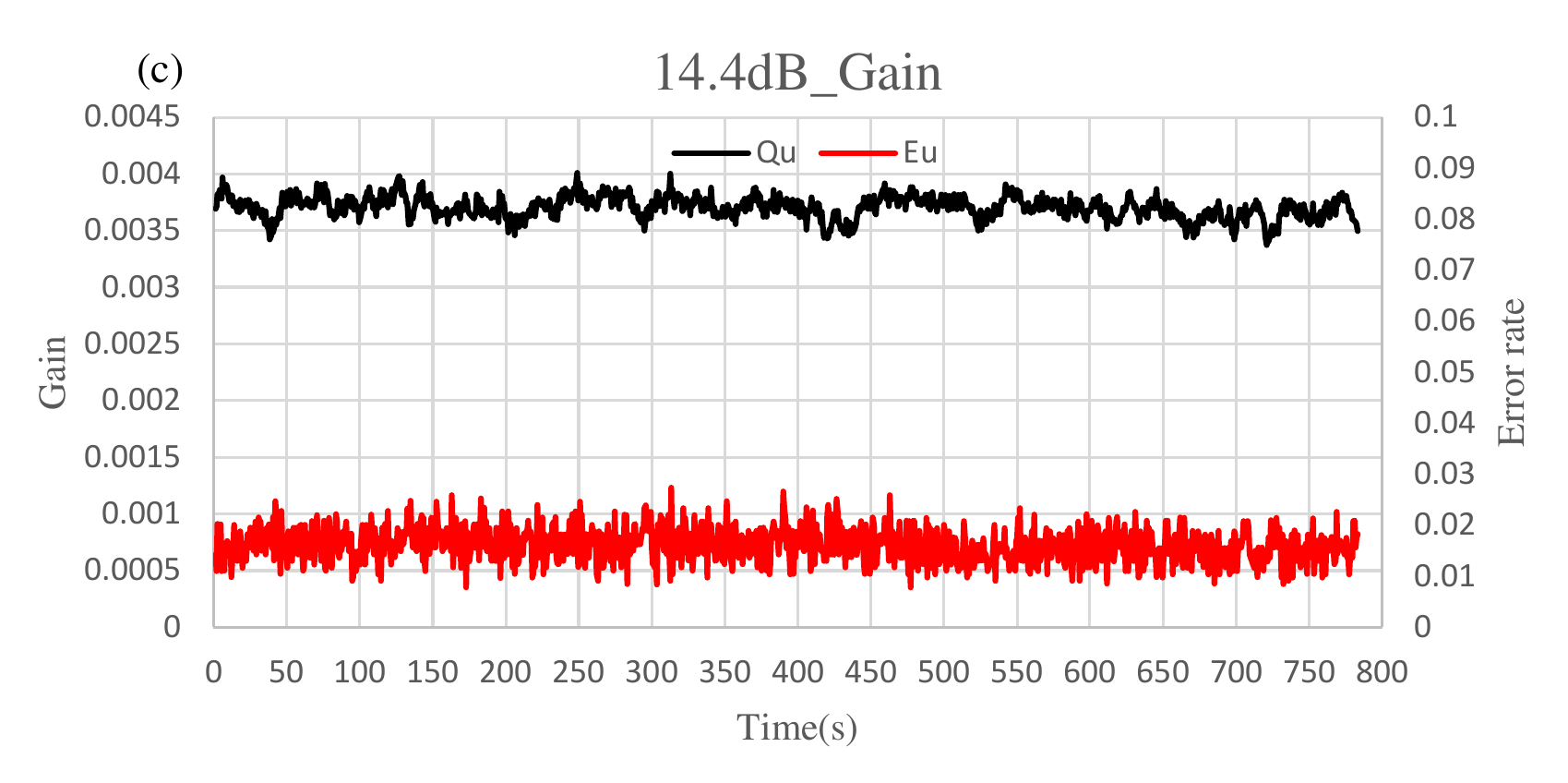}\includegraphics[width=6.5cm]{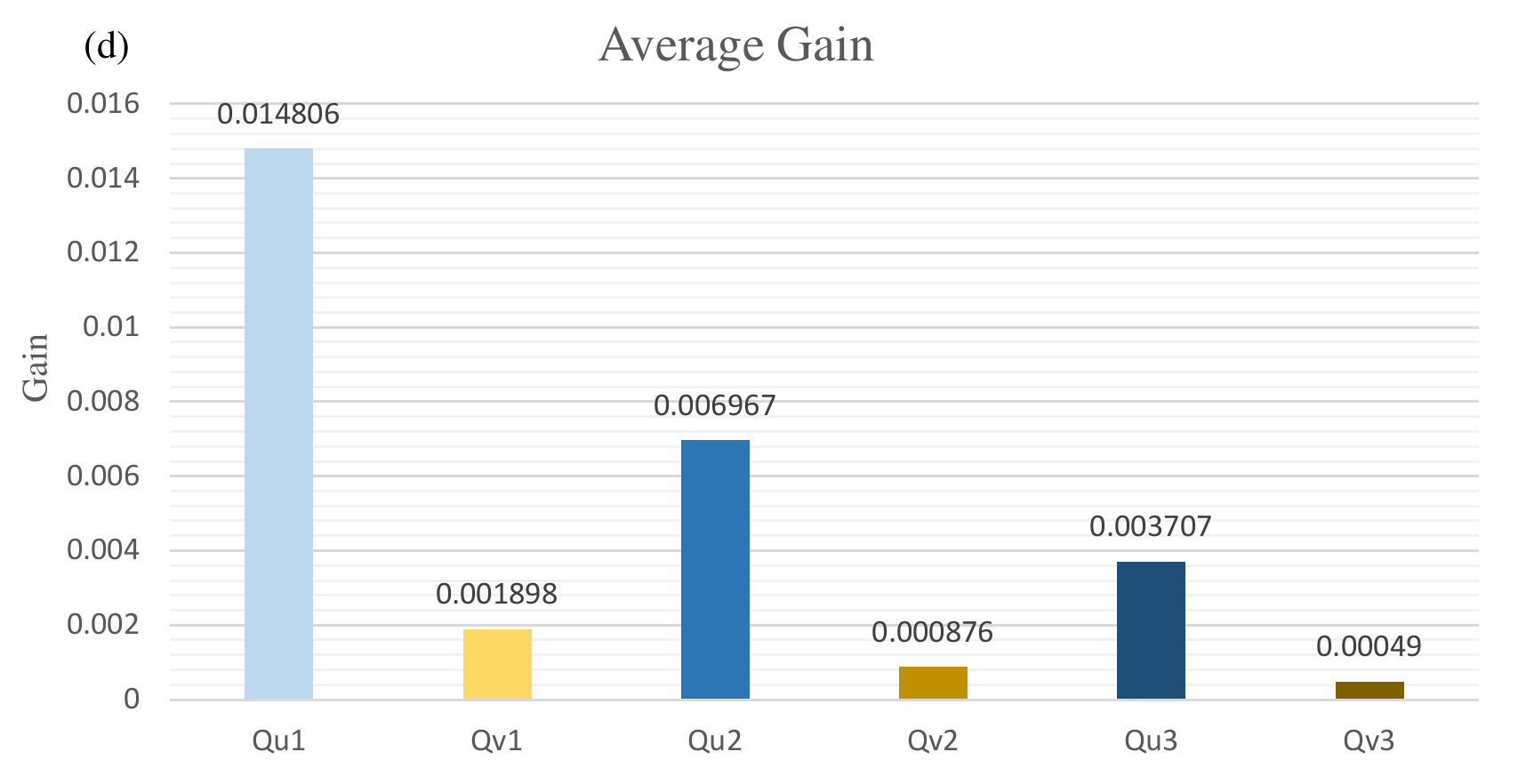}
    \caption{Experimental results.(a), (b) and (c) Real-time gain of signal state and error rate under different channel attenuation. (a)10.22dB, duration 526.2s. (b)12.36dB, duration 500.1s. (c)14.4dB,duration 783.3s. (d) Average gain of signal state and decoy state under three different water channels.}
\end{figure}

For the other two scenarios, the gain of the signal state and decoy state both suffer from different degrees of decay due of the gradually increased channel attenuation. In 12.36dB scenario, the $Q_u$ has reduced to $6.97\times10^{-3}$ and $Q_v$ has reduced to $8.76\times10^{-4}$. In the 14.40dB scenario, the values of these two parameters drop to $3.71\times10^{-3}$ and $4.90\times10^{-4}$.However, the bit error rates of these scenarios have not gone too worse.The bit error rates of signal states are 1.64\% for the 12.36dB and 1.62\% for the 14.40dB cases. Considering the well-preserved polarization state after transmission delivers the main contribution to this result. The average final key rates for these two scenarios are 826.4 bits/s and 404.6 bits/s, respectively.

\begin{table}[]
	\renewcommand\arraystretch{1.5}
	\caption{Crucial parameters of the experiments}\label{Table 2}
    \centering  
    \scriptsize
	\begin{tabular}{|p{45pt}<{\centering}|p{30pt}<{\centering}|p{30pt}<{\centering}|p{30pt}<{\centering}|p{40pt}<{\centering}|p{30pt}<{\centering}|p{80pt}<{\centering}|}
		\hline		
		$Attenuation$&$Q_u$&$E_u$&$Q_v$&$Q_1$&$E_1$&$Final key rate(bits/s)$\\
		\hline
		10.22 dB&1.48×10$^{-2}$&$1.21\%$&1.89×10$^{-3}$&4.84×10$^{-3}$&$1.18\%$&$3535.7$\\
		\cline{1-7}\cline{2-7} \cline{3-7} \cline{4-7} \cline{5-7} \cline{6-7} \cline{7-7} 
		12.36 dB&6.97×10$^{-3}$&$1.64\%$&8.76×10$^{-4}$&1.94×10$^{-3}$&$0.71\%$&$826.4$\\
		\cline{1-7}\cline{2-7} \cline{3-7} \cline{4-7} \cline{5-7} \cline{6-7} \cline{7-7}  
		14.40 dB&3.70×10$^{-3}$&$1.62\%$&4.90×10$^{-4}$&1.01×10$^{-3}$&$1.52\%$&$404.6$\\
		\cline{1-7}\cline{2-7} \cline{3-7} \cline{4-7} \cline{5-7} \cline{6-7} \cline{7-7} 
		16.35 dB&1.47×10$^{-3}$&$2.20\%$&2.17×10$^{-4}$&5.51×10$^{-4}$&$1.91\%$&$245.6$\\
		\cline{1-7}\cline{2-7} \cline{3-7} \cline{4-7} \cline{5-7} \cline{6-7} \cline{7-7} 
		\hline
	\end{tabular}
	\normalsize
\end{table} 

In order to verify that our self-designed decoy-state laser module can calibrate the trigger intensity of the signal state and decoy state freely and independently, we readjust the average photon number per pulse of the signal state to 0.7, whereas the decoy state remains unchanged. The system is then be tested under 16.35dB channel attenuation for executing QKD procedure. During 939s experimental time, the gain of signal state and the bit error rate are $1.47\times10^{-3}$ and 2.20\% , the final key rate reaches 245.6 bits /s. 

The final key rates of the four experiments show that our system has yet to reach the maximal performance. Therefore, we conduct numerical simulations based on our experimental setup. The primary indicator of a QKD system is the secure key rate. Here we calculate the secure key rate by Eq. (1) given in Ref. \cite{ma2005practical}, i.e., 

\begin{equation}
R_{SKR}\geq q\{-Q_{u}f(E_{u})H_{2}(E_{u})+Q_{1}[1-H_{1}(e_{1})]\}
\end{equation}

Here $q$ is the sifting rate of the QKD process(the value in our simulation is 0.5 ), $Q_1$,and $e_1$ represent the gain and the error rate of single photon. $f(x)$ is the bidirectional error correction efficiency. $H_2(x)$ represents the binary Shannon information function. The simulated results of the relation between secure key rates and water channel distances are shown in Fig. 5. The demonstrated results are based on three sets of channel attenuation, and the attenuation coefficients of the Jerlov type \uppercase\expandafter{\romannumeral1},\uppercase\expandafter{\romannumeral2} and \uppercase\expandafter{\romannumeral3} are 0.018, 0.13 and 0.29.

\begin{figure}[hbtp]
    \centering
    \includegraphics[width=9cm]{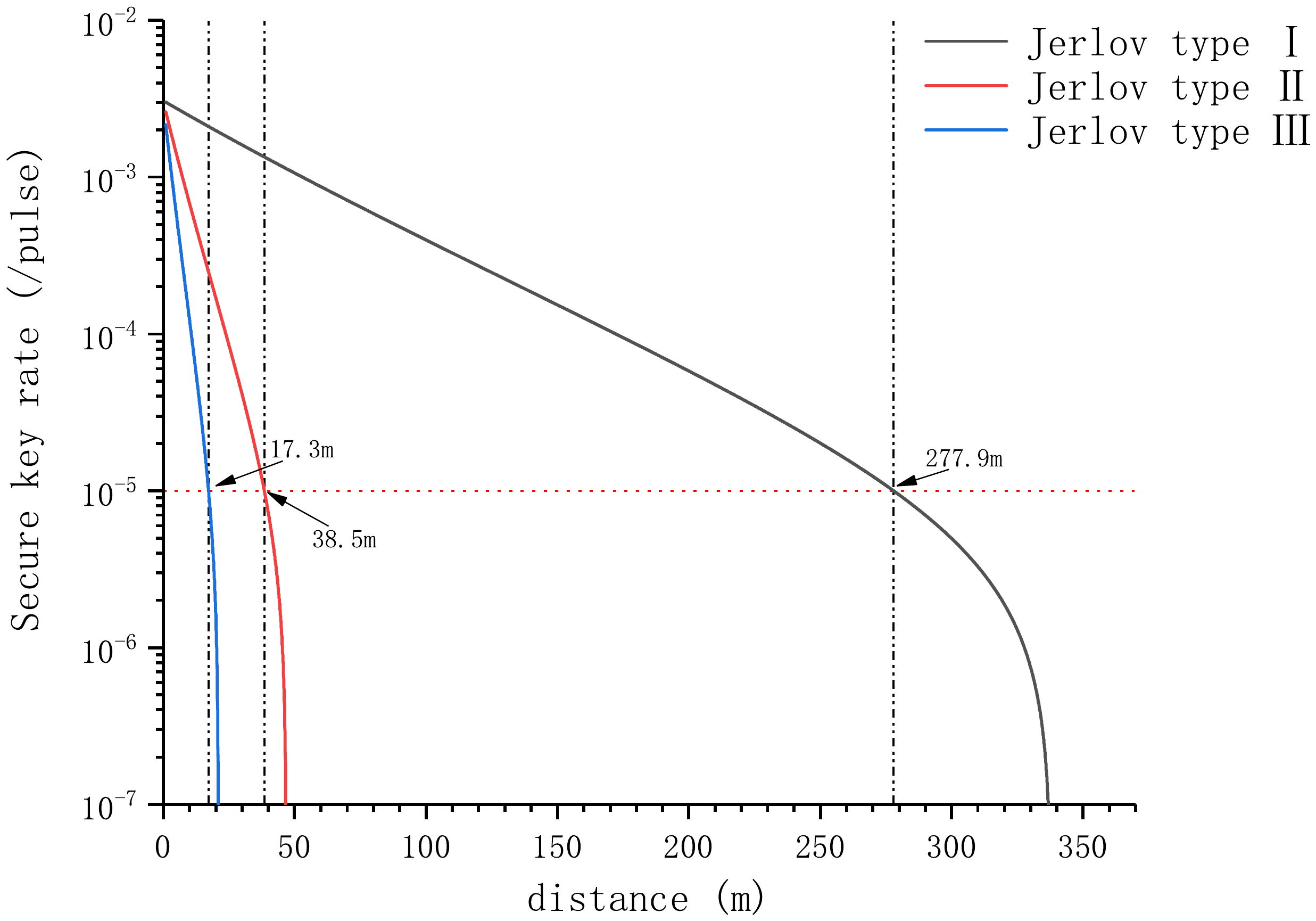}
    \caption{Simulation results.Relation between secure key rate(per pulse) and water channel distance under different channel attenuation. }
    \label{fig:my_label}
\end{figure}

In the 16.35dB channel loss experiment, the channel attenuation is equal to a 209m Jerlov type \uppercase\expandafter{\romannumeral1} water channel. According to the simulation results shown in Fig. 5, if we wish to accomplish further distance at the cost of lowering final key rates, the system can achieve a final key rate of $1\times10^{-5}$ per pulse (around 30 bits/s) at a distance of 277.9m, in which case the channel loss is 21.7 dB. Considering 11.1dB of optical attenuation at Bob end, the system can still work under 32.8dB of total attenuation.

The overall experimental results confirm that the performance of our self-designed decoy-state laser module is good in the decoy-state QKD system. The experiments under different channel attenuation show that the new system can successfully execute decoy-state BB84 QKD protocol. Compared with our previous work, the new system ditches four redundant laser modules and fully redesigns the system structure. More importantly, based on our self-designed decoy-state laser module, the underwater decoy-state QKD system is significantly more compact in space.

\section{ Conclusion and discussion}
We successfully devised a decoy-state laser module, based on which, a new compact underwater decoy-state QKD system was developed and tested. The pulse intensity of signal state and decoy state can be separately adjusted and calibrated on the self-designed decoy-state laser module. Based on the laser module, the overall dimension of the transmitter/receiver in the newly developed system is as small as $64.5cm\times23.5cm\times20cm$. During the QKD experiments, the performance of our self-designed decoy-state laser module is proved well , and a final key rate of 3.54Kbps is obtained with an average bit error rate of 1.18\% under a 10.22dB channel attenuation.By adjusting the ratio of the average photon number per pulse between the signal state and the decoy state from 8:1 to 7:1, the final key rate of 245.6bps is obtained with an average bit error rate of 1.91\% under a 16.35dB channel attenuation, which is equal to 209m Jerlov type \uppercase\expandafter{\romannumeral1} seawater. Furthermore, we demonstrated that our system can still generate secure keys in the 277.9m Jerlov type \uppercase\expandafter{\romannumeral1} water channel, in which the channel attenuation is up to 21.7dB. The system is completely controlled by the FPGA and will be ready for the sea field test once suitably packaged. 

There are some practical issues that we will be faced in next phase of experiments. The most important of which is the alignment between Alice and Bob. Currently we have spent time developing and testing the control algorithm of oscillating mirror, which can be a part of the acquisition pointing tracking (APT) system. With the help of the APT system, we can steer the light beam from Alice into the Bob’s receiving window when the deviation of their positions is in a reasonable range. Moreover, the automatic posture perception and the adjustment system when facing with relatively large position misalignment will also be considered in the future studies.  

\begin{backmatter}
\bmsection{Funding}
This work was supported by the National Natural Science Foundation of China  (Grants No. 61701464 and No. 61575180 ) and the Fundamental Research Funds for the Central Universities (Grants No. 202165008).

\bmsection{Acknowledgments}
The authors thank Xiangnian Shang Ya Xiao and Xiaobing Hei for their help in preparing the paper.

\bmsection{Disclosures}
The authors declare no conflicts of interest. 

\bmsection{Data Availability Statement}
Data underlying the results presented in this paper are not publicly available at this time but may be obtained from the authors upon reasonable request

\end{backmatter}

\bibliography{sample.bib}

\end{document}